\documentclass[aps,prd,nofootinbib,amsmath,superscriptaddress]{revtex4}

\usepackage{bm}
\usepackage{graphicx}

\newcommand{\be}{\begin{equation}}
\newcommand{\ee}{\end{equation}}
\newcommand{\bea}{\begin{eqnarray}}
\newcommand{\eea}{\end{eqnarray}}

\newcommand{\bfp}{\mbox{\boldmath $p$}}

\newcommand{\bfP}{\mbox{\boldmath $P$}}

\newcommand{\bfy}{\mbox{\boldmath $y$}}

\newcommand{\bfT}{\mbox{\boldmath $T$}}

\newcommand{\Lup}{\Lambda^\uparrow}

\newcommand{\ua}{\uparrow}
\newcommand{\da}{\downarrow}

\begin{document}

\title{The Polarising Fragmentation Function and the \\ $\Lambda$ polarisation 
in $e^+e^-$ processes}

\author{M.~Anselmino}
\email{mauro.anselmino@to.infn.it}
\affiliation{Dipartimento di Fisica, Universit\`a di Torino,
             Via P.~Giuria 1, I-10125 Torino, Italy}
\affiliation{INFN, Sezione di Torino, Via P.~Giuria 1, I-10125 Torino, Italy}
\affiliation{Department of Physics, Indian Institute of Technology Bombay,
Mumbai-400076, India}
\author{R.~Kishore}
\email{raj.theps@gmail.com}
\affiliation{Department of Physics, Indian Institute of Technology Bombay,
Mumbai-400076, India}
\author{A.~Mukherjee}
\email{asmita@phy.iitb.ac.in}
\affiliation{Department of Physics, Indian Institute of Technology Bombay,
Mumbai-400076, India}

\date{\today}

\begin{abstract}
The surprising polarisation of $\Lambda$s and other hyperons measured in 
many unpolarised hadronic processes, $p\,N \to \Lup \, X$, has been a long 
standing challenge for QCD phenomenology. One possible explanation was suggested, 
related to non perturbative properties of the quark hadronisation process, and 
encoded in the so-called Polarising Fragmentation Function (PFF). Recent Belle 
data have shown a non zero $\Lambda$ polarisation also in unpolarised $e^+e^-$ 
processes, $e^+ e^- \to \Lambda \, X$ and $e^+ e^- \to \Lambda \, h\, X$. We 
consider the single inclusive case and the role of the PFFs. Adopting a 
simplified kinematics it is shown how they can originate a polarisation 
$P_\Lambda \not= 0$ and give explicit expressions for it in terms of the PFFs. 
Although the Belle data do not allow yet, in our approach, an extraction of the 
PFFs, some clear predictions are given, suggesting crucial measurements, and 
estimates of $P_\Lambda$ are computed, in qualitative agreement with the Belle 
data. 
          
\end{abstract}


\maketitle

\section{\label{intro}Introduction}
The polarisation of $\Lambda$ hyperons inclusively produced in the high energy 
interactions of unpolarised hadrons, $p \, N \to \Lup \, X$~\cite{Bunce:1976yb,
Heller:1978ty}, is a major challenge for QCD theoretical interpretations 
since many years. A large amount of data is available~\cite{Heller:1996pg}, due 
to the fact that the weak decay of the $\Lambda$ allows an easy measurement of 
its polarisation $P_\Lambda$. No definite explanation of the origin of $P_\Lambda$, 
in a QCD framework, is convincingly available. In the usual application of 
perturbative QCD and collinear factorisation, the elementary interactions among 
unpolarised partons cannot produce any significant final state quark
polarisation~\cite{Kane:1978nd}. 

Non perturbative QCD features have been invoked. In Ref.~\cite{Anselmino:2000vs}
and in Ref.~\cite{Anselmino:2001js} Transverse Momentum Dependent (TMD) effects 
in the fragmentation process were introduced, adopting a TMD factorisation scheme, 
respectively in proton-proton ($p\,p$) and lepton-proton ($\ell \, p$) interactions. 
In Ref.~\cite{Zhou:2008fb} collinear higher-twist quark-gluon-antiquark correlations 
in the nucleon were considered for $\ell\,p$ and $p \, p$ processes, while in 
Refs.~\cite{Koike:2017fxr} and~\cite{Gamberg:2018fwy} a complete twist-3 
collinear fragmentation contribution to polarised hyperon production in 
unpolarised hadronic and $e^+e^-$ collisions has been presented. The TMD effects 
in the quark fragmentation of Refs.~\cite{Anselmino:2000vs,Anselmino:2001js} were 
encoded in the so called Polarising Fragmentation Function (PFF), introduced 
and defined in Ref.~\cite{Mulders:1995dh}.  

Very recently new data on the polarisation of $\Lambda$ hyperons produced in 
unpolarised $e^+e^-$ annihilation processes, $e^+e^- \to \Lambda \, h \, X$ 
and $e^+e^- \to ({\rm jet}) \, \Lambda \, X$, have been published by the Belle 
Collaboration~\cite{Guan:2018ckx}, showing a non zero value of $P_\Lambda$.
Prompted by these data and the simplicity of the process which only depends
on fragmentation functions, we consider in this paper the role of the transverse 
momentum dependent PFF in generating the $\Lambda$ polarisation. A general 
theoretical discussion of two hadron inclusive production in $e^+e^-$ 
interactions can be found in Ref.~\cite{Boer:1997mf}. 

Our aim is that of understanding the physical mechanism through which the 
particular correlation between the momentum of the fragmenting quark, the 
momentum of the final $\Lambda$ and its polarisation - described by the PFF -
can build up the $\Lambda$ polarisation. To do so we will adopt a very simple
kinematical configuration which allows analytical computations and a direct 
visualisation of the process. It allows to understand specific features of 
$P_\Lambda$, which are true beyond the approximate kinematics and provide 
genuine testable predictions. Although the actual data of the Belle 
Collaboration, as it will be explained in Section~\ref{section3}, do not 
allow a precise direct comparison with our computation of $P_\Lambda$, our 
estimates result in good qualitative agreement with the data.

\section{\label{section2}Formalism and simple analytical results for $P_\Lambda$} 
\begin{figure}[]
\begin{center}
\includegraphics[width=15.truecm,angle=0]{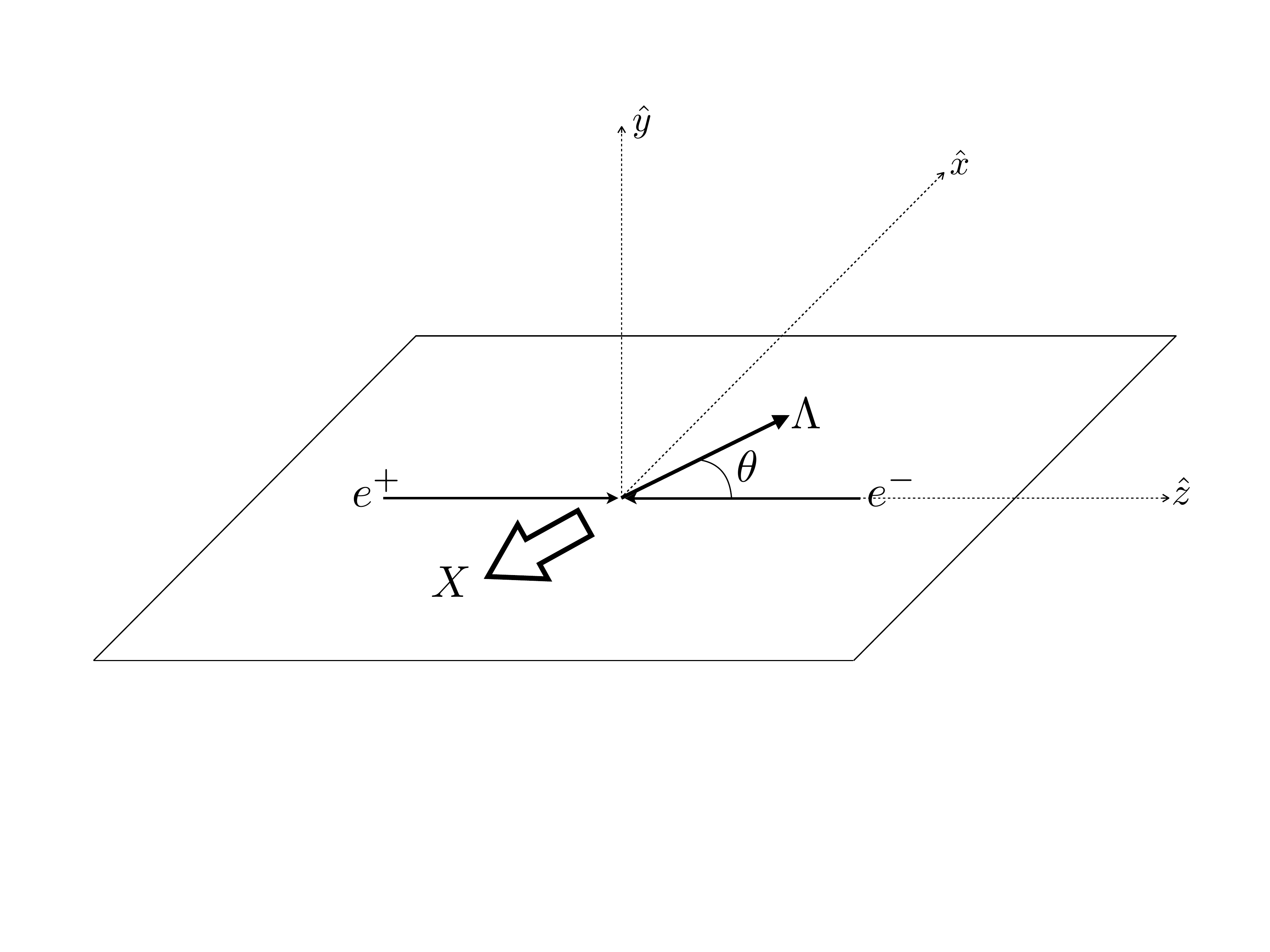}
\vskip -80pt 
\caption{Definition of the c.m. reference frame for the process $e^+e^- \to 
\Lambda \, X$}
\label{fig:Lambdaplane}
\end{center}
\end{figure}

We consider the process $e^+e^- \to \Lambda(\bfP)  \, X$ in the c.m. reference 
frame defined in Fig.~1, with:     
\be
\bfp_{e^+} \equiv \bfp = \frac {\sqrt s}{2} (0, \, 0, \, 1) \quad\quad\quad  
\bfp_{e^-} = -\bfp = \frac {\sqrt s}{2} (0, \, 0, \, -1) \quad\quad\quad
\bfp_\Lambda = p_\Lambda (\sin\theta, \, 0, \, \cos\theta) \>, \label{frame}
\ee
where masses have been neglected. The initial leptons are unpolarised, while 
we consider the spin polarisation vector $\bfP$ of the $\Lambda$ hyperon. 
Notice that, by parity invariance, the dependence of the cross section on 
$\bfP$ must be of the form $\bfP \cdot \bfp \times \bfp_\Lambda \sim \sin\theta$ 
and $\bfP$ can only be perpendicular to the production plane. 

The $\Lambda$ production, at leading order, goes via the subprocess 
$e^+e^- \to q \, \bar q$, with the subsequent fragmentation of a quark into the
$\Lambda$, such that 
\be
\bfp_\Lambda = z \, \bfp_q + \bfp_\perp \> \label{plambda}
\ee 
where 
\be
\bfp_q \cdot \bfp_\perp = 0 \quad\quad\quad  
p_\Lambda^2 = \frac{z^2s}{4} + p_\perp^2 \>.
\ee

The fragmentation function of an unpolarised quark into a spin 1/2 $\Lambda$ 
hyperon with polarisation vector along the $\uparrow = \bfP$ or $\downarrow = 
-\bfP$ directions can be written as 
\be
D_{\Lambda^{\uparrow, \downarrow}/q}(z, \bfp_\perp) = 
\frac12 D_{\Lambda/q}(z, p_\perp) 
\pm \frac12 \, \Delta^ND_{\Lup/q}(z, p_\perp) \, \bfP \cdot \hat{\bfp}_q
\times \hat{\bfp}_\perp
\ee
where $D_{\Lambda/q}(z, p_\perp)$ is the unpolarised Transverse Momentum 
Dependent Fragmentation Function (TMD-FF) and $\Delta^ND_{\Lup/q}(z, p_\perp)$ 
is the Polarising Fragmentation Function (PFF): it encodes basic features of the 
quark hadronisation process and describes the number density of spin 1/2 polarised 
hadrons ($\Lambda$ hyperons in this case) resulting from the fragmentation of 
an unpolarised quark. The angular dependence, $\bfP \cdot \hat{\bfp}_q
\times \hat{\bfp}_\perp$, is dictated by parity invariance. 

The cross section for the production of a $\Lambda$ hyperon with spin
polarisation vector $\ua$ or $\da$ in $e^+e^-$ annihilation, can be written, 
assuming TMD factorisation at leading order~\cite{Boglione:2017jlh}, as a 
convolution of the elementary annihilation cross section ($d\hat\sigma^q$) 
and the fragmentation function:
\be
d\sigma^{\ua,\da} = \sum_q e_q^2 \> d\hat\sigma^q \otimes 
D_{\Lambda^{\uparrow, \downarrow}/q} \>, \label{conv}
\ee 
where $q = u, \bar u, d, \bar d, s, \bar s$ quarks. 
The unpolarised cross section is given by
\be
d\sigma^{\ua} + d\sigma^{\da} = \sum_q e_q^2 \> d\hat\sigma^q \otimes 
D_{\Lambda/q} \label{unp} \>,
\ee    
the cross section difference by
\be
d\sigma^{\ua} - d\sigma^{\da} = \sum_q e_q^2 \> d\hat\sigma^q \otimes 
\Delta^ND_{\Lup/q}(z, p_\perp) \, \bfP \cdot \hat{\bfp}_q
\times \hat{\bfp}_\perp \>, \label{diff}
\ee
and the $\Lambda$ polarisation in the $\bfP$ direction is given by
\be
P_\Lambda = \frac {d\sigma^{\ua} - d\sigma^{\da}}
{d\sigma^{\ua} + d\sigma^{\da}} \>\cdot \label{pol} 
\ee

The convolutions in Eqs.~(\ref{conv})-(\ref{diff}) should take into account 
all possible momenta $\bfp_q$ of the fragmenting quark and all possible 
values of $\bfp_\perp$ such that $\bfp_\Lambda = z\,\bfp_q + \bfp_\perp$.  
Then, in general, the quark momentum $\bfp_q$ has also components outside 
the $xz$ $\Lambda$ production plane. However, the main contribution to a 
polarisation perpendicular to the $xz$ plane is originated, because of the 
cross product $\hat{\bfp}_q \times \hat{\bfp}_\perp$, by momenta in that plane. 
When computing the polarisation we then consider the simple kinematical 
configuration given in Fig. 2, in which, at fixed values of $z$ and $p_\perp$, 
there are two vectors $\bfp_\perp$ contributing to $\bfp_\Lambda$: 
\be
\bfp_{q_{1,2}} = \frac {\sqrt s}{2} (\sin\theta_{q_{1,2}}, \, 0, \, 
\cos\theta_{q_{1,2}}) \quad\quad\quad
\bfp_\perp^{(1,2)} = p_\perp (\pm \cos\theta_{q_{1,2}}, \,0, \, 
\mp \sin\theta_{q_{1,2}})    
\ee
where we have already imposed the condition $\bfp_q \cdot \bfp_\perp = 0$.
\begin{figure}[]
\begin{center}
\includegraphics[width=15.truecm,angle=0]{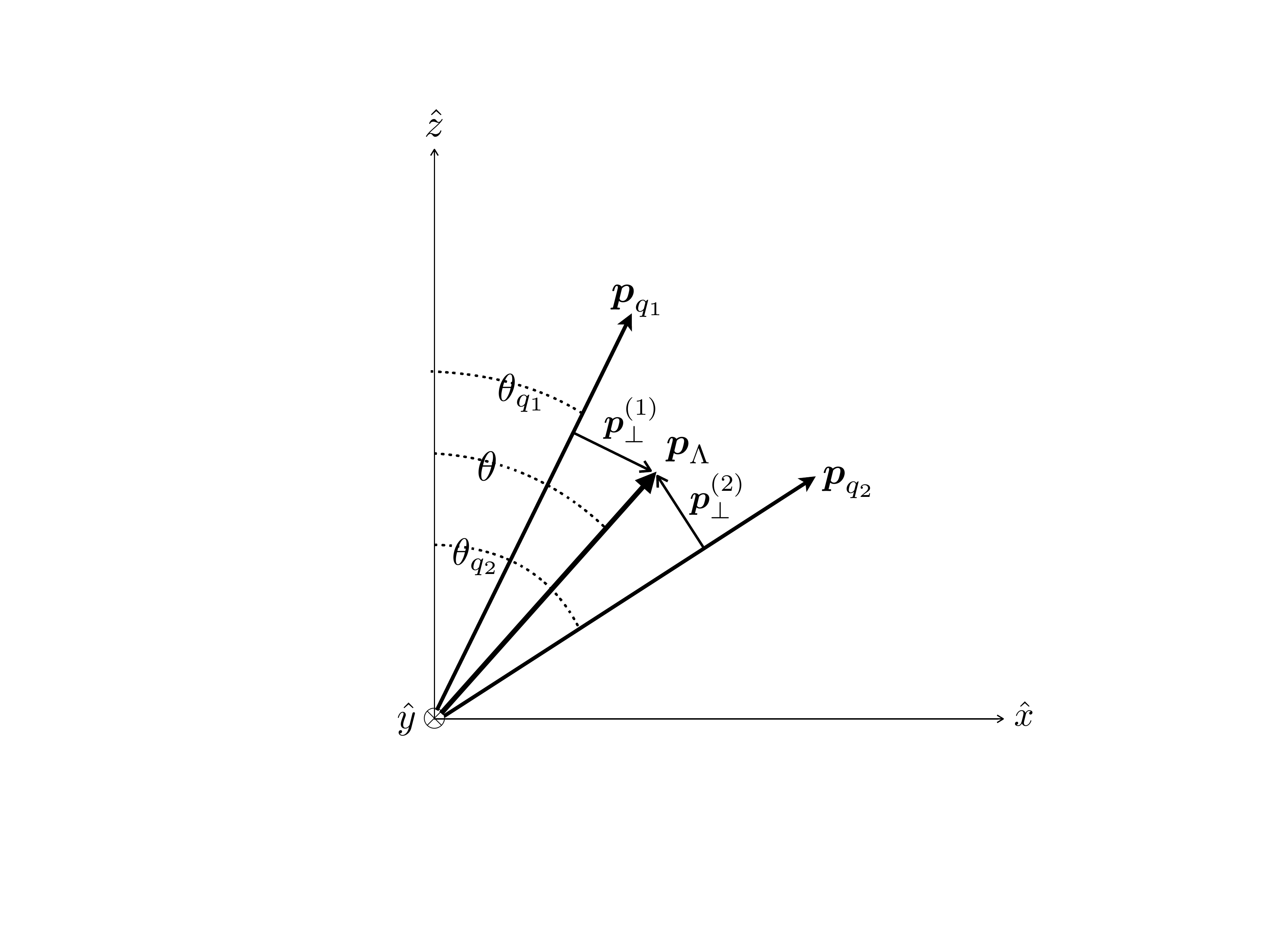}
\vskip -72pt 
\caption{Simple 2-dimensional kinematical configuration. The $\hat y$-axis is 
entering the $xz$ plane.}
\label{fig:2d}
\end{center}
\end{figure}

We must also impose the condition~(\ref{plambda}): 
\bea
p_\Lambda \sin\theta &=& z \frac {\sqrt s}{2} \sin\theta_{q_{1,2}} \pm
p_\perp \cos\theta_{q_{1,2}} \\
p_\Lambda \cos\theta &=& z \frac {\sqrt s}{2} \cos\theta_{q_{1,2}} \mp
p_\perp \sin\theta_{q_{1,2}}
\eea
with the solutions
\bea
\cos\theta_{q_{1,2}} &=& \frac{1}{\sqrt{1 + \frac 
{\displaystyle 4p_\perp^2}{\displaystyle z^2s}}}
\left( \cos\theta \pm \frac {2p_\perp}{z \sqrt s}\sin\theta \right) 
\label{cos12}\\
\sin\theta_{q_{1,2}} &=& \frac {1}{\sqrt{1 + \frac 
{\displaystyle 4p_\perp^2}{\displaystyle z^2s}}}
\left( \sin\theta \mp \frac {2p_\perp}{z \sqrt s}\cos\theta \right) \cdot
\label{sin12}
\eea

As we said, the only polarisation allowed by parity invariance is orthogonal 
to the $xz$ production plane, that is $\bfP = \hat\bfy$. Notice that, in 
Eq.~(\ref{diff}), according to our simple kinematical configuration of Fig.~2,
we have:    
\be
\hat{\bfp}_{q_1} \times \hat{\bfp}_\perp^{(1)} = + \hat\bfy  \quad\quad\quad
\hat{\bfp}_{q_2} \times \hat{\bfp}_\perp^{(2)} = - \hat\bfy \>.
\ee
The convolution in Eq.~(\ref{diff}) then reads:
\be
\frac {d\sigma^{\ua}}{dz \, d(\cos\theta)} -
\frac {d\sigma^{\da}}{dz \, d(\cos\theta)} =
\sum_q e_q^2 \int \! p_\perp \, dp_\perp \left[ 
\frac {d\hat\sigma^{q}}{d(\cos\theta_{q_1})} -
\frac {d\hat\sigma^{q}}{d(\cos\theta_{q_2})} \right]
\> \Delta^ND_{\Lup/q}(z, p_\perp) \>, \label{main} 
\ee
where the elementary $e^+e^- \to q\,\bar q$ cross section is given by
\be
\frac {d\hat\sigma^{q}}{d\cos(\theta_q)} = \frac{3\pi\alpha^2}{2s}
(1 + \cos^2\theta_q) \label{qqbar}
\ee
and the expressions of $\cos\theta_{q_{1,2}}$ are given in Eq.~(\ref{cos12}). 

By using Eqs.~(\ref{qqbar}) and (\ref{cos12}) in Eq.~(\ref{main}) one has
a very simple result:
\bea
\frac {d\sigma^{\ua}}{dz \, d(\cos\theta)} -
\frac {d\sigma^{\da}}{dz \, d(\cos\theta)} &=&
\frac{3\pi\alpha^2}{2s} \sum_q e_q^2 \int \! p_\perp \, dp_\perp 
( \cos^2\theta_{q_1} - \cos^2\theta_{q_2} ) 
\> \Delta^ND_{\Lup/q}(z, p_\perp) \label{main1} \\
&=& \frac{3\pi\alpha^2}{2s} \sum_q e_q^2 \int \! p_\perp \, dp_\perp 
\frac {4 \, z \, p_\perp \sqrt s}{z^2s + 4 \, p_\perp^2} \> \sin(2\theta)
\> \Delta^ND_{\Lup/q}(z, p_\perp) \>. \label{main2}
\eea
Some more comments on the physical interpretation of this expression and 
the approximations involved will be made in the next Section. 

We have now to compute the unpolarised cross section, appearing in the 
denominator of Eq.~(\ref{pol}). As, differently from the polarised case, now
all components of $\bfp_q$ contribute equally to the $\Lambda$ production, 
and not only those in the $xz$ plane, we write the convolution~(\ref{unp}) as:    
\bea
\frac {d\sigma}{dz \, d(\cos\theta)} &=&
\sum_q e_q^2 \int \! 2\pi \, p_\perp \, dp_\perp
\frac {d\hat\sigma^{q}}{d(\cos\theta)} 
\> D_{\Lambda/q}(z, p_\perp) \label{den} \\
&=& 
\sum_q e_q^2 \> \frac {d\hat\sigma^{q}}{d(\cos\theta)} 
\> D_{\Lambda/q}(z) \label{den2}
\eea
where, essentially, we have assumed $d\hat\sigma^q/d(\cos\theta_q) \simeq 
d\hat\sigma^q/d(\cos\theta)$ and used the relation $\int d^2\bfp_\perp 
D_{\Lambda/q}(z, p_\perp) = D_{\Lambda/q}(z)$. Eq.~(\ref{den2}) is the 
usual expression for the cross section in the collinear partonic configuration. 

By collecting Eqs.~(\ref{pol}) and (\ref{qqbar})--(\ref{den2}) 
we have simple expressions for the $\Lambda$ polarisation along the 
$\hat\bfy$ direction:
\be
P_\Lambda (z, \cos\theta) = \frac{\displaystyle \sum_q e_q^2
\int \! p_\perp \, dp_\perp 
\frac {4 \, z \, p_\perp \sqrt s}{z^2s + 4 \, p_\perp^2}
\> \Delta^ND_{\Lup/q}(z, p_\perp)}
{\displaystyle \sum_q e_q^2  
\> D_{\Lambda/q}(z)}
\> \frac {\sin(2\theta)}{1 + \cos^2\theta} \label{polzt}
\ee   
\be
P_\Lambda (z, p_\perp) = \frac{\displaystyle \sum_q e_q^2 \>
\frac {4 \, z \, p_\perp \sqrt s}{z^2s + 4 \, p_\perp^2} 
\> \Delta^ND_{\Lup/q}(z, p_\perp)}
{\displaystyle \sum_q e_q^2 \> 2\pi \> D_{\Lambda/q}(z, p_\perp)}
\> \frac{\displaystyle \int \! d(\cos\theta)\> \sin(2\theta)}
{\displaystyle \int \! d(\cos\theta)\> (1 + \cos^2\theta)} \label{polzp}
\ee 

\be
P_\Lambda (z, p_\perp, \cos\theta) = \frac{\displaystyle \sum_q e_q^2
\> \frac {4 \, z \, p_\perp \sqrt s}{z^2s + 4 \, p_\perp^2} 
\> \Delta^ND_{\Lup/q}(z, p_\perp)}
{\displaystyle \sum_q e_q^2 \> 2\pi \> D_{\Lambda/q}(z, p_\perp)}
\> \frac {\sin(2\theta)}{1 + \cos^2\theta} \>\cdot \label{polzpt}
\ee 

In the single inclusive $\Lambda$ production process that we are considering, 
the only observables are $z$ and $\cos\theta$, while the values of $p_\perp$ 
are integrated. We have also given Eqs.~(\ref{polzp}) and~(\ref{polzpt}) as
they might allow some comparison with the data of Ref.~\cite{Guan:2018ckx}.

\section{\label{section3}Comments, suggested measurements and some predictions}

Before trying to give some estimates of the polarisation a few comments are in 
order, which illustrate the meaning and validity of the results obtained in the 
previous Section. 

\begin{itemize}
\item
Our simple 2-dimensional kinematical configuration shows clearly the mechanism 
which, thanks to the $\bfP \cdot \bfp_q \times \bfp_\perp$ correlation of the PFF, 
builds up the $\Lambda$ polarisation. As illustrated in Fig.~\ref{fig:2d},
at any fixed values of $z$ and $p_\perp = |\bfp_\perp|$ there are two possible 
vectors $\bfp_q$ leading to the same $\bfp_\Lambda$. These two vectors 
lead to opposite vectors $\bfp_q \times \bfp_\perp = \bfp_q \times \bfp_\Lambda$
and then to opposite values of the polarisation along the $\bfP = \hat\bfy$
direction. However, in the convolution of Eqs.~(\ref {diff}) and (\ref{main}),
to each of them there corresponds a different value of the scattering angle 
$\theta$ and then of the elementary cross section (\ref{qqbar}). This leads 
to a clear physical interpretation of Eqs.~(\ref {main1}) and (\ref{main2}). 

\item
Notice also that the annihilation cross section is symmetric around $\theta = 
\pi/2$, where $\cos^2\theta_{q_{1}} = \cos^2\theta_{q_{2}}$, Eq.~(\ref {cos12}); 
thus, the polarisation vanishes at $\theta = \pi/2$, 
reflecting the nature of the $e^+e^- \to q \, \bar q$ partonic interaction 
at leading order. In addition, as we already remarked, the polarisation must 
vanish at $\theta = 0$ and $\theta = \pi$ for parity invariance: then one 
understands the $\sin(2\theta)$ behaviour of $P_\Lambda$.

\item
The true kinematical implementation of our mechanism should take into account
a 3-dimensional configuration of $\bfp_q$ around $\bfp_\Lambda$. Then, always 
at fixed values of $z$ and $p_\perp$, we would have contributions from other
pairs of vectors $\bfp_{q_{1}}$ and $\bfp_{q_{2}}$ out of the $xz$ plane. 
Such vectors contribute (oppositely) to $P_\Lambda$ only with their components 
in the $xz$ plane, which is smaller than $p_\perp$. In addition, for them the 
difference between $\theta_{q_{1}}$ and $\theta_{q_{2}}$ is also smaller. 
We estimate that their total contribution to $P_\Lambda$ would not change 
the value obtained in our limited 2-dimensional kinematical model by more 
than a factor 2. Our numerical computations will actually underestimate 
the value of the polarisation. 

\item
The conclusions of the first two paragraphs of this Section are valid also 
in full 3-dimensional kinematics; the first one simply illustrates the PFF 
mechanism, while the second one is based on the symmetry properties, at 
leading order, of the elementary cross section, proportional to 
$(1 + \cos^2\theta)$. Then, our result
\be 
P_\Lambda \sim \sin(2\theta) \label{sin2t}
\ee
is a genuine prediction of the way in which a PFF would build up the 
$\Lambda$ polarisation; a prediction which could and should be easily 
tested experimentally. 

\item
We have given explicit expressions of $P_\Lambda (z, p_\perp, \cos\theta)$,
and some of their integrated forms, in terms of the unknown polarising 
fragmentation function $\Delta^ND_{\Lup/q}(z, p_\perp)$. The $\theta$ dependence,
as we stressed in the previous paragraph, is instead well defined. Notice that
we also find 
\be 
P_\Lambda = {\cal O} (p_\perp/\sqrt s),
\ee  
meaning that the polarisation in the single inclusive process 
$e^+e^- \to \Lambda \, X$ is a higher twist effect. It agrees with the 
observation of a vanishing $\Lambda$ transverse polarisation in $e^+e^-$ at 
the $Z^0$ pole by the OPAL Collaboration~\cite{Ackerstaff:1997nh}.
\end{itemize}

In the single inclusive process that we are considering, one only observes the 
final $\Lambda$ and its polarisation, given by Eq.~(\ref{polzt}). Such a process
allows to exploit the fact that one knows the direction of $\bfP_\Lambda$,
perpendicular to the production plane. In addition, we have assumed that also 
the elementary dynamics takes predominantly place in the same plane. While the 
$\cos\theta$ dependence of the polarisation is definitely fixed by the PFF
mechanism, independently of the simplified kinematics, its magnitude, 
$P_\Lambda (z, \cos\theta)$, depends on our assumption and on the functional 
form of the PFF $\Delta^ND_{\Lup/q}(z, p_\perp)$. If abundant and precise data
were available for $P_\Lambda$ in the $e^+e^- \to \Lambda \, X$ process, one 
could try to parametrise the PFF, and, performing a best fit of the experimental 
points, extract information on the PFF.  

Unfortunately, the available experimental information is not exactly what we 
would like to have in order to perform such a procedure. In 
Ref.~\cite{Guan:2018ckx} the inclusive $\Lambda$ polarisation is measured along 
the direction ${\hat\bfT} \times {\hat\bfp_\Lambda}$, where ${\hat\bfT}$ is the 
thrust axis, which, apart from unavoidable experimental inaccuracies, coincides
with ${\hat\bfp}_q$. What actually the Belle Collaboration measure is the 
polarisation in the process $e^+e^- \to {\rm jet} \, \Lambda \, X$, which, in 
general, needs not be perpendicular to the $\Lambda$ production plane.
Their polarisation is presented as a function of $p_\perp$ and $z$. Presumably, 
data have been collected over a wide range of the angle $\theta$, although this 
information cannot be found in the paper. 

However, we can attempt a very qualitative comparison with the Belle data, by 
exploiting Eq.~(\ref{polzpt}), which also gives, with some approximations, 
the polarisation along $\bfp_q \times \bfp_\Lambda$ as a function of $p_\perp$, $z$ 
and $\cos\theta$. Similarly, if we knew the range of $\theta$ covered by the 
experiment, we could use Eq.~(\ref{polzp}). 

\begin{figure}[]
\begin{center}
\includegraphics[width=12.truecm,angle=0]{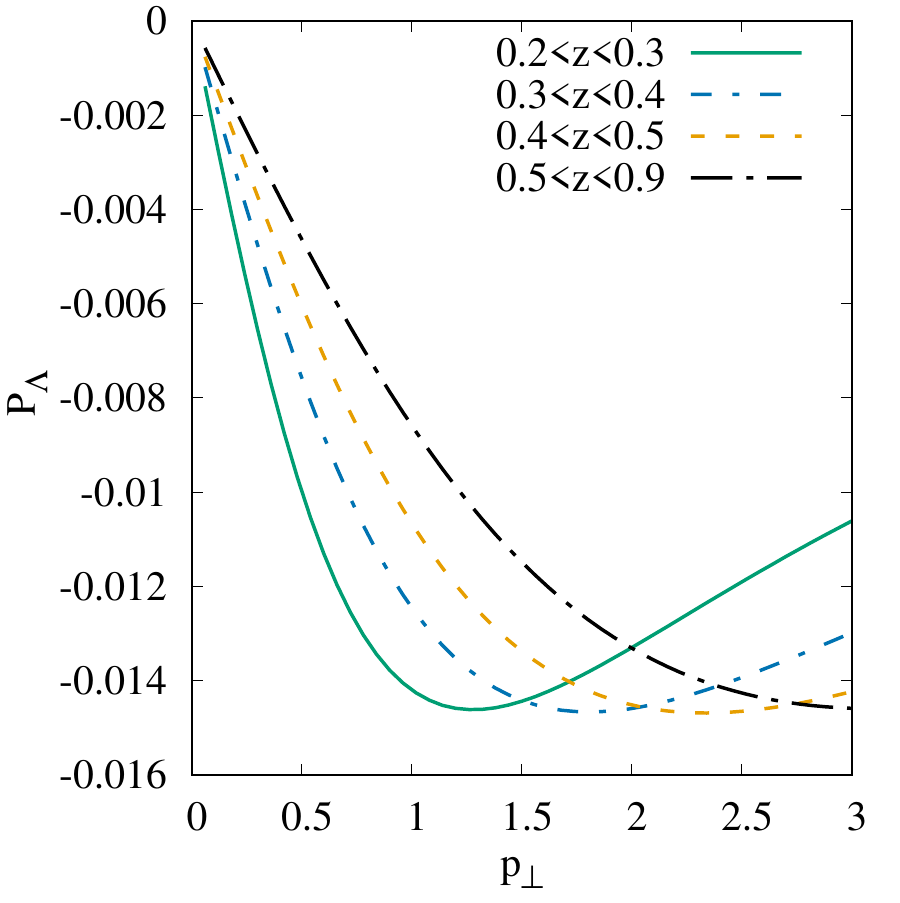}
\caption{Plots of $P_\Lambda (p_\perp, \theta =\pi/3))$ vs.~$p_\perp$, 
computed according to Eq.~(\ref{polzpt}). For each bin of $z$ values the 
numerator and denominator of Eq.~(\ref{polzpt}) have been accordingly 
integrated. The PFFs are given in Eq.~(\ref{20percent}) and the FFs 
$D_{\Lambda/q}(z)$ are from Ref.~\cite{deFlorian:1997zj}.}
\label{fig:lp_pt}
\end{center}
\end{figure}

\begin{figure}[]
\begin{center}
\includegraphics[width=12.truecm,angle=0]{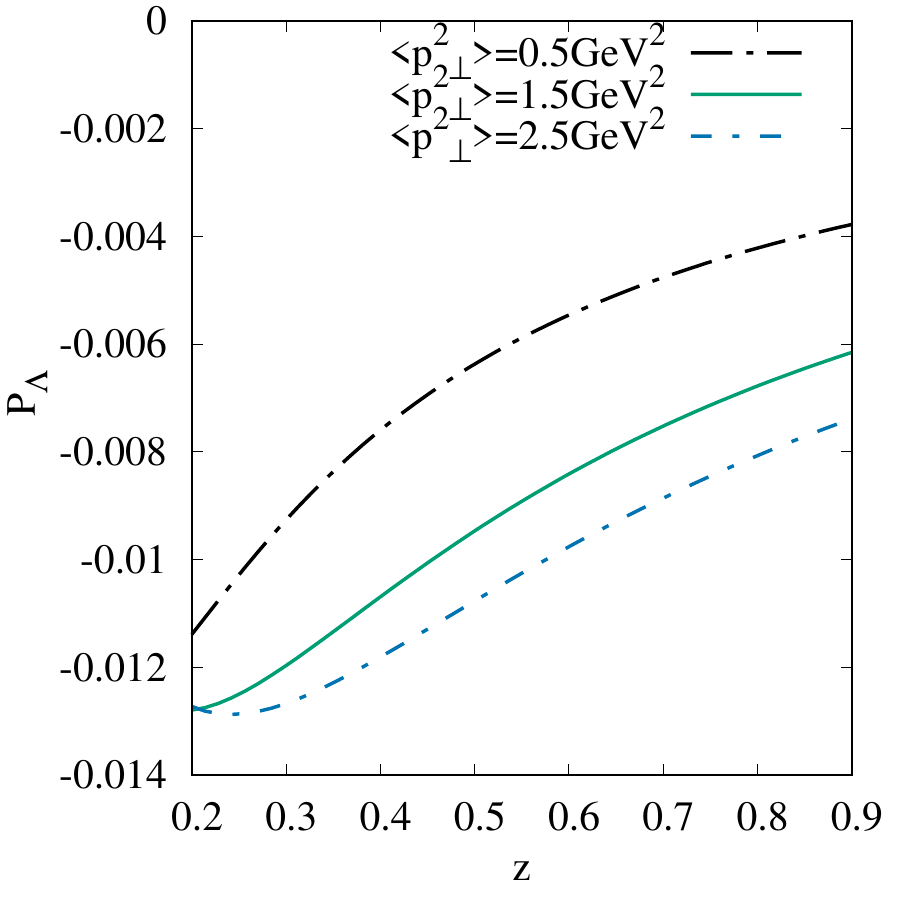}
\caption{Plots of $P_\Lambda (z, \theta =\pi/3))$ vs.~$z$, 
computed according to Eq.~(\ref{polzt}), for different values of 
$\langle p^2_{\perp}\rangle$. We have integrated numerator and 
denominator of Eq.~(\ref{polzt}) in the range $0 \leq p_\perp \leq 3$ GeV.}
\label{fig:lp_z}
\end{center}
\end{figure}
 
In order to obtain simple numerical estimates of $P_\Lambda$ we assume the 
PFFs, $\Delta^ND_{\Lambda^{\uparrow}/q}(z,p_{\perp})$, to be $20\%$ of the 
corresponding unpolarised TMD-FFs, $D_{\Lambda/q}(z,p_{\perp})$. In 
Ref.~\cite{Anselmino:2001js} the PFFs have been parameterised in terms of 
the unpolarised FFs and have been used to compute the $\Lambda$ polarisation 
in $p\,p \to \Lambda \, X$ processes. From the best fit parameters it is found 
that the overall normalisation factor for the $u$ and $d$ quark PFFs has to be 
negative whereas for $s$ quarks it has to be positive. Motivated by this, we 
choose 
\begin{eqnarray}
\Delta^ND_{\Lambda^{\uparrow}/u}(z,p_{\perp})=-0.2\times D_{\Lambda/u}(z,p_{\perp})
\nonumber\\
\Delta^ND_{\Lambda^{\uparrow}/d}(z,p_{\perp})=-0.2\times D_{\Lambda/d}(z,p_{\perp})
\label{20percent}\\
\Delta^ND_{\Lambda^{\uparrow}/s}(z,p_{\perp})=0.2\times D_{\Lambda/s}(z,p_{\perp})
\>. \nonumber 
\end{eqnarray} 

As usual, for the unpolarised TMD-FFs, we consider factorised $z$ and $p_\perp$
dependences, with a Gaussian parameterisation for the transverse momentum 
dependent part: 
\begin{eqnarray}
D_{\Lambda/q}(z,p_{\perp}) = 
D_{\Lambda/q}(z) \> \frac{1}{\pi\,\langle p^2_{\perp}\rangle} \>
e^{\displaystyle -\frac{p^2_{\perp}}{\langle p^2_{\perp}\rangle}} \>,
\label{tmdff}
\end{eqnarray}
where $q=u,d,s$ and $D_{\Lambda/q}(z)$ is the collinear FF. The width of the 
Gaussian, $\langle p^2_{\perp}\rangle$, is a free parameter which could be 
extracted by fitting data from experiments.

In order to further simplify our order of magnitude estimates, we consider the 
LO collinear FF from Ref.~\cite{deFlorian:1997zj}, where it has been assumed,
following simple $SU_f(3)$ symmetry arguments, that all the light flavour 
quarks fragment to $\Lambda$ with equal probability, i.e.,
\begin{eqnarray}{\label{cffs}}
D_{\Lambda/u}=D_{\Lambda/d}=D_{\Lambda/s} \equiv D_{\Lambda/q}\>,
\end{eqnarray}
while the antiquark fragmentation into a $\Lambda$ are taken to be negligible,
$D_{\Lambda/\bar q} = 0$. 

In the plots given in Fig.~\ref{fig:lp_pt} we show $P_\Lambda$ as a function 
of $p_\perp$, computed according to Eq.~(\ref{polzpt}), for different bins of 
$z$, similar to those of the Belle data~\cite{Guan:2018ckx}, and fixing 
$\theta = \pi/3$. For each plot we have integrated the numerator and denominator 
of Eq.~(\ref{polzpt}) over $z$ according to the corresponding bin range.
The LO collinear FFs $D_{\Lambda/q}(z)$ for light quarks have been taken from  
Ref.~\cite{deFlorian:1997zj}. Notice that, in this case, as a consequence  
of Eqs.~(\ref{tmdff}) and~(\ref{20percent}), the flavour independent Gaussian 
widths $\langle p^2_{\perp}\rangle$ do not affect the value of 
$P_{\Lambda}(p_\perp, \theta)$.
  
As we commented before, the actual comparison between our computation of 
$P_\Lambda$ and the Belle data can only be considered at a qualitative level.
Our definition of $z$, Eq.~(\ref{plambda}), differs from the value of $z_\Lambda 
= 2E_\Lambda/\sqrt s$, used in Ref.~\cite{Guan:2018ckx}, by terms of the order 
$ p_\perp^2/(sz^2)$. Moreover, our estimates of $P_\Lambda$ undervalue it. 
The signs of the PFFs have been assumed to be the same as those obtained in 
fitting the $\Lambda$ polarisation in $p\,p \to \Lambda \, X$ processes~\cite{Anselmino:2001js}. Despite all this,
when comparing with Fig.~1 of Ref.~\cite{Guan:2018ckx}, 
the qualitative agreement is remarkable, with negative values of 
$P_\Lambda(p_\perp)$ of the order of a few percents or less. Notice, however, 
that the sign of $P_\Lambda$ depends on the value of the production angle
$\theta$ and changes at $\theta = \pi/2$.   

In Fig.~\ref{fig:lp_z} we plot $P_\Lambda$ as a function of $z$ according to 
Eq.~(\ref{polzt}), for different values of $\langle p_\perp^2 \rangle$, 
fixing $\theta = \pi/3$ and integrating numerator and denominator over 
$p_\perp$ between 0 and 3 GeV. Increasing the upper integration limit has 
negligible effects. Notice that, in this case, again as a consequence  
of Eqs.~(\ref{tmdff}) and~(\ref{20percent}), the collinear FFs $D_{\Lambda/q}(z)$
cancel out in the ratio giving $P_{\Lambda}(z, \theta)$. 

Again, with our choice of the PFFs, Eq.~(\ref{20percent}), we obtain, at the
chosen $\theta$ angle, negative values of $P_{\Lambda}(z)$, of the order of 
1\%, in qualitative agreement with the Belle data~\cite{Guan:2018ckx}.     
  
\section{\label{section4}Conclusions}   
Motivated by new data on the polarisation of $\Lambda$ hyperons, measured 
by the Belle Collaboration in unpolarised $e^+e^-$ annihilation 
processes~\cite{Guan:2018ckx}, we have investigated the role of the Polarising 
Fragmentation Functions~\cite{Anselmino:2000vs,Anselmino:2001js,Mulders:1995dh} 
in generating the polarisation. These new data are particularly interesting as 
they refer to a process in which $P_\Lambda$ can only depend on the properties 
of the quark fragmentation, without the complication of convolutions with 
partonic distributions, such as in the case of $p \, N \to \Lup \, X$ and 
$\ell \, N \to \Lup \, X$ processes.

We have adopted a simplified 2-dimensional kinematics, which might lead to an 
underestimate of $P_\Lambda$, but has the merit of visualising the partonic 
process and the role of the PFFs. In addition, in our LO TMD factorisation
scheme, we obtain a general clear prediction for the $\theta$ dependence of 
$P_\Lambda$, Eq.~(\ref{sin2t}), which could be easily tested.     

A precise comparison of the results of our approach with the few Belle data 
is not possible yet, because of the nature of the Belle data, which do not 
refer to a fully inclusive process, and to the uncertainty on the $\Lambda$ 
production angle $\theta$, not discussed in Ref.~\cite{Guan:2018ckx}. 
However, by making simple assumptions on the PFFs and the collinear FF, and 
following indications on the sign of the PFFs obtained by fitting the $\Lambda$ 
polarisation in $p \, p \to \Lup \, X$ processes~\cite{Anselmino:2000vs}, we 
have given some estimates of the expected values of $P_\Lambda(p_\perp)$ and 
$P_\Lambda(z)$ in the same $p_\perp$ and $z$ kinematical regions covered by 
the Belle measurements, fixing a particular value of $\theta$. Such estimates 
show a negative and small $\Lambda$ polarisation, in agreement with the Belle 
data.        

\acknowledgments 
We are grateful to W. Vogelsang for sharing with us his code for the 
unpolarised FFs, $D_{\Lambda/q}(z)$. We thank M.~Boglione, U. D'Alesio,
F. Murgia and A. Prokudin for useful comments and discussions.

%
\end{document}